\documentclass[aps,pra,twocolumn,showpacs]{revtex4}
\usepackage{graphicx}
\usepackage{bbm} % cool fonts: try $\mathbbm 1$
\usepackage{amsmath}
\usepackage{amssymb}
\usepackage{psfrag}
\newcommand{\bra}[1]{\langle #1|}
\newcommand{\ket}[1]{| #1 \rangle}
\newcommand{\braket}[2]{\langle #1|#2 \rangle}

\newcommand{\expi}[1]{ {\rm e}^{{\rm i} #1}}

\newcommand{\cre}[1]{\hat{#1}^\dagger}  %Creation operator
\newcommand{\des}[1]{\hat{#1}}  %Destruction operator

\newcommand{\op}[1]{\hat{#1}} %operator
\newcommand{\APSimage}[4]{
\begin{figure}[t]
   \begin{center}
      \includegraphics[angle = 0, width = #2]{#1}
   \end{center}
\caption{\label{#4}#3}
\end{figure}
}
\begin{document}

\title{Ultra-large Rydberg dimers in optical lattices}

\author{B. Vaucher}
 \email{benoit.vaucher@merton.ox.ac.uk}
 \homepage{http://www.physics.ox.ac.uk/qubit}
\author{S.~J.~Thwaite}
\author{D.~Jaksch}
\affiliation{Clarendon Laboratory, University of
Oxford, Parks Road, OX1 3PU, United Kingdom}

\date{\today}

\begin{abstract}

We investigate the dynamics of Rydberg electrons excited
from the ground state of ultracold atoms trapped in an
optical lattice. We first consider a lattice comprising an
array of double-well potentials, where each double well is
occupied by two ultracold atoms. We demonstrate the
existence of molecular states with equilibrium distances of
the order of experimentally attainable inter-well spacings
and binding energies of the order of $10^3\, \mathrm{GHz}$.
We also consider the situation whereby ground-state atoms
trapped in an optical lattice are collectively excited to
Rydberg levels, such that the charge-density distributions
of neighboring atoms overlap. We compute the hopping rate
and interaction matrix elements between highly-excited
electrons separated by distances comparable to typical
lattice spacings. Such systems have tunable interaction
parameters and a temperature $\sim 10^{4}$ times smaller
than the Fermi temperature, making them potentially
attractive for the study and simulation of strongly
correlated electronic systems.
\end{abstract}

\pacs{37.10.Jk, 34.20.Cf, 32.80.Ee}

%37.10.Jk,34.20.Cf,32.80.Ee
\maketitle

\section{Introduction}

Recent advances in the trapping and manipulation of
ultracold atomic gases have provided experimentalists with
the ability to coherently control large numbers of atoms.
Two areas that have become the focus of experimental
efforts of late are the use of ultracold atoms for the
formation and manipulation of molecules
\cite{Hutson2006,Thalhammer2006} and the creation and
manipulation of Rydberg atoms in optical
lattices~\cite{Knuffman2007}. In this paper we study
whether combining these areas might lead to the production
of diatomic molecules whose nuclear position is fixed by an
optical lattice (see Figs.~\ref{fig:setup}a--b). In
particular, we examine the properties of ultralarge
dimers with equilibrium distances of the order of typical
lattice spacings and binding energies of the order of
$10^3\, \mathrm{GHz}$. We also investigate the prospect of
using systems of interacting Rydberg atoms to simulate
Fermi systems.

\APSimage{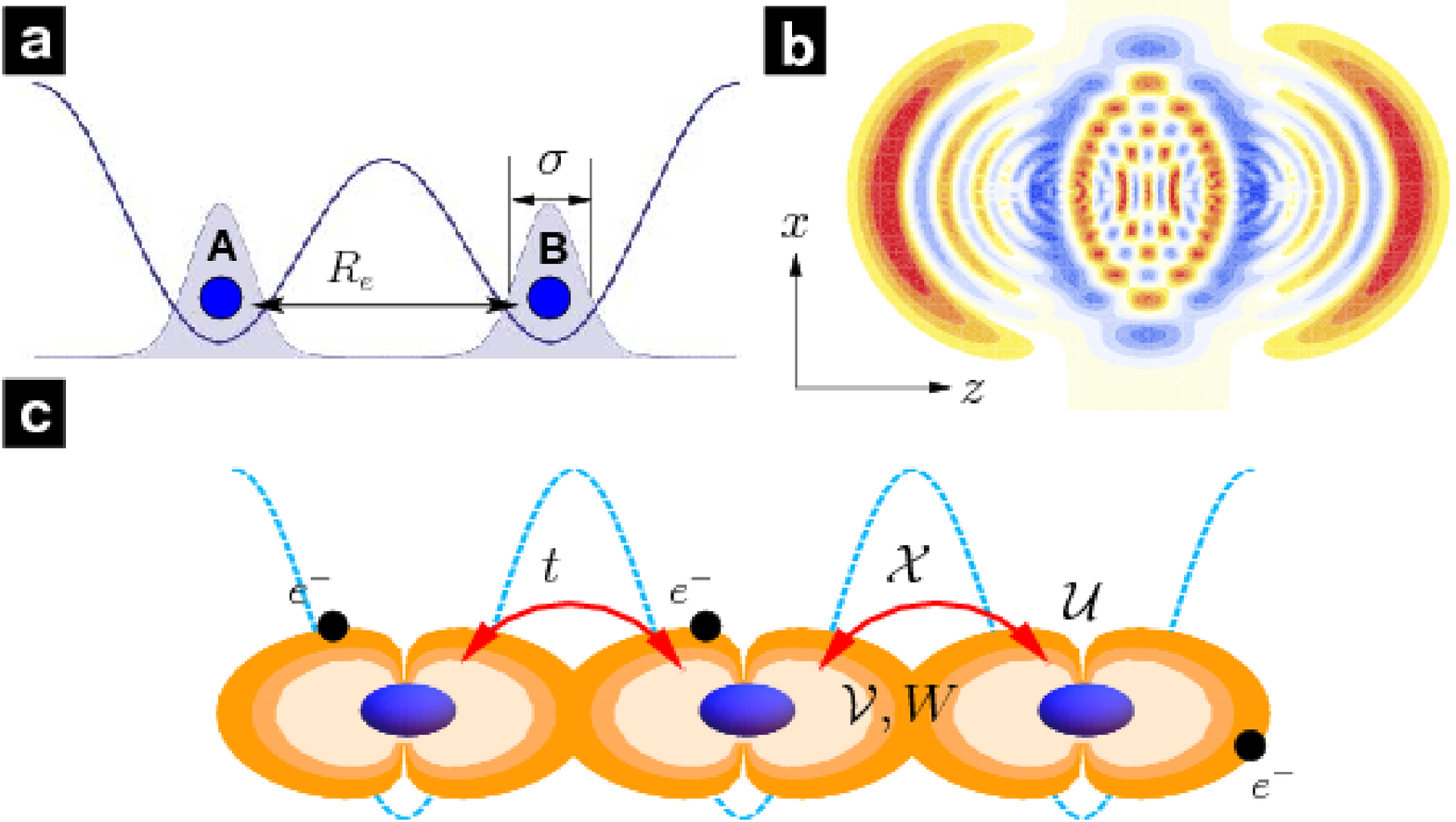}{8cm}{(color online) An optical lattice
with pairs of atoms well separated from each other is
initialized (a). A laser pulse transfers each pair of atoms to
a molecular state with a very large internuclear distance.
(b) Density plot of a typical diatomic molecular
wavefunction on the $x$-$z$ plane ($y=0$) in the relative
coordinates of two electrons in highly excited $np_z$
states. (c) The outer electron of each ground-state atom
trapped in an optical lattice is transferred into a Rydberg
state, such that the charge-density distributions between
neighboring atoms overlap. The electron hopping rate $t$
becomes non-zero, and the interactions between electrons
can be described by the parameters $\mathcal{U}$ (on-site
interaction), $\mathcal{V}$, $\mathcal{W}$ and
$\mathcal{X}$ (off-site interactions). }{fig:setup}

Molecules have a far richer energy structure than atoms,
and can also have stronger long-range interactions, a
feature which offers new possibilities for quantum control
(see e.g.~\cite{DeMille2002}). However, cooling molecules
is notably difficult, since the absence of closed
electronic transitions prevents the use of standard laser
cooling procedures. An attractive approach to producing
translationally cold molecules is thus to form them from
pre-cooled atoms by way of photoassociation or magnetic
resonance techniques.  In recent years several classes of
ultracold molecules have been predicted and produced,
including Feshbach molecules, Efimov trimers, and the
famous `trilobite' molecules, which are composed of a
highly excited Rydberg atom interacting with another atom
in its ground state~\cite{Greene2000, Khus2002}. The
existence of long-range molecules composed of two Rydberg
atoms stabilized via dipole-dipole interactions has also
been predicted~\cite{Farooqi2003,
Boisseau2002,Stanojevic2008}.
% 20 nanoK = 400 Hz=10^2 Hz
% In Boiseau, De~ 3e-2 cm-1 ~> 9 x10^2 MHz =10^8 Hz
These ultracold long-range Rydberg molecules have binding
energies of the order of several hundred MHz (a factor of
$\sim 10^6$ greater than the typical temperature of
ultracold atoms, but weak by molecular standards), a
lifetime expected to be similar to that of Rydberg atoms,
and equilibrium distances of the order of $10^4 \, a_0$
(with $a_0$ the Bohr radius)~\cite{Boisseau2002}. Recent
advances in high-resolution microscopy techniques may offer
new possibilities to study the spatial structure of such
molecules~\cite{Gericke2008}. Theoretical treatments of
Rydberg molecules to date have primarily dealt with the
regime in which the internuclear separation is greater than
the Le Roy radius. In this regime the overlap between the
atomic charge-density distributions is vanishingly
small~\cite{Boisseau2002,Farooqi2003}. In contrast to these
works we consider the regime where the internuclear
distance of the molecular dimer is comparable to typical
lattice spacings, but smaller than the Le Roy radius, in
which case the overlap of the charge distributions and
effects such as the exchange interaction must be taken into
account.

The simulation of condensed-matter systems is another area
to which optical lattices are uniquely suited. Optical
lattices have extremely flexible geometries, and by a
suitable choice of laser configuration, any desired lattice
structure can be created \cite{Petsas1994,Saers2008}.
Fermions loaded into an optical lattice can be used as a
model of electrons in a solid, with the significant
advantage of having a periodic potential that is defectless
and fully customizable~\cite{Lewenstein2006}. However, the
ratio between the currently attainable temperature $T$ of
fermionic atoms trapped in an optical lattice and the Fermi
temperature $T_F$ of the system is $T/T_F\sim 0.25$,
preventing the clean observation of such interesting
phenomena as the BCS transition or the emergence of certain
types of anti-ferromagnetic
order~\cite{Kohl2006,Idziaszek2001, Lewenstein2006}. In the
final section of this paper we consider the possibility of
collectively transferring a population of ground-state
atoms trapped in an optical lattice to a highly excited
Rydberg level, such that the charge-density distributions
of neighboring atoms overlap~(see Fig.~\ref{fig:setup}c).
We extend the model developed in the first section to
compute the hopping rate and interaction parameters between
the highly excited electrons, and determine the dependence
of these quantities on the initial lattice spacing. While
the implementation of these systems is experimentally
challenging and beyond the scope of this paper
\cite{Cubel2005}, we note that experimental temperatures
far smaller than the Fermi temperature $(T/T_F \sim
10^{-4})$ might be readily attainable, making the
realisation of such systems a promising new approach to the
quantum simulation of interacting fermions. % fermionic matter

The paper is organized as follows. In Sec.~\ref{sec:mol} we
present the model that we subsequently use to investigate
the existence of attractive molecular potentials with
equilibrium distances below the Le Roy radius. We
approximate the lifetime of these molecules and calculate
their equilibrium internuclear distance for different
values of the principal quantum number of the electrons. In
Sec.~\ref{sec:simul} we use the model introduced in the
first section to compute the hopping rate and interaction
parameters of the Rydberg electrons and discuss the
possibility of using these setups to simulate
condensed-matter systems. We conclude in
Sec.~\ref{sec:conclusion}.

%=--------------------------------------------

\section{Molecular potentials of ultra-large Rydberg dimers\label{sec:mol}}

In this section we present a method of evaluating the
energy of highly excited diatomic molecular states whose
equilibrium distance $R_e$ is below the Le Roy radius. We
develop an efficient method of computing one- and
two-center molecular integrals over electronic orbitals
with large principal quantum numbers $n$ and calculate the
depth and equilibrium position of molecular potentials
accessible from the ground state via photoassociation for
values of $n$ up to $n=35$. We estimate the radiative
lifetime of the highly excited molecular potentials.

Quantum chemistry methods commonly used within the Le Roy
radius typically become computationally expensive when
dealing with Rydberg molecules due to the size of the basis
set required to describe the spatially diffuse Rydberg
orbitals. In order to obtain qualitative results in this
computationally challenging regime we take a simple
wavefunction \emph{ansatz} inspired by the Heitler-London
treatment of the hydrogen molecule. Due to the simplistic
form of the molecular wavefunctions used we do not expect
the method presented to produce quantitatively accurate
results; rather, our aim is to obtain qualitatively correct
results and insights within a regime where common methods
of calculating molecular potentials fail or become
computationally costly.

%=--------------------------------------------

\subsection{Model}
We consider an optical lattice potential forming an array
of double-wells separated from each other by sufficiently
high optical barriers that each double well can be
considered as an isolated system~\cite{Sebby-Strabley2006}.
We assume that each double well initially contains two
$^{87}\mathrm{Rb}$ atoms in their ground state, and that
the potential is sufficiently deep that the Wannier
function associated with each atom is well localized in one
half of the well. Because alkali atoms have only a single
valence electron, their energy levels are described by the
same quantum numbers as those of the hydrogen atom. If the
ground-state atoms are excited to Rydberg levels with
$n>30$ by applying a laser pulse to the system, the charge
distributions of atoms in the same double well will
overlap, and under certain conditions a stable molecular
state will be formed. We aim to study the basic properties
of these molecules and explore whether they may be produced
in an optical lattice with experimentally realistic
parameters. Since the temperature of the system is very
low, the velocity of the electrons -- even in highly
excited states -- is much greater than that of the trapped
ions, and the electrons respond almost instantaneously to
displacements of the ions~\cite{Boisseau2002}.
Consequently, we will calculate molecular potentials in the
Born-Oppenheimer (BO) approximation
~\cite{Atkins2005,Slater1974}.

Neglecting interactions between atoms belonging to
different double wells, the energy of a homonuclear
molecule formed by the atoms in a double well is given
within the BO approximation by the eigenvalues of the
Hamiltonian
\begin{align}
H(R)&=
-\frac{\hbar^2}{2m_e}\left( \nabla_{\mathbf{r}_1}^2 + \nabla_{\mathbf{r}_2}^2\right)\nonumber\\
&+ j_0 \left( \frac{1}{R}+\frac{1}{r_{12}}
-\frac{1}{r_{1A}}-\frac{1}{r_{1B}}-\frac{1}{r_{2A}}-\frac{1}{r_{2B}}
\right),\label{diatomicH}
\end{align}
where $\mathbf{r}_i$ ($i=1,2$) is the position of the
$i^\mathrm{th}$ electron; $r_{i\xi}$ ($\xi=A,B$) is the
distance between the $i^\mathrm{th}$ electron and atomic
center $\xi$; $r_{12}$ is the distance between the two
electrons; $j_0 = e^2/(4\pi\varepsilon_0)$, where
$\varepsilon_0$ is the permittivity of free space and $e$
the elementary unit of charge; and $R$ is the distance
between the nuclei. In the BO approximation $R$ is treated
as a classical variable upon which the eigenvalues and
eigenfunctions of the Hamiltonian depend parametrically. 
Following a similar approach to that of
Ref.~\cite{Stanojevic2006,Stanojevic2008}, we approximate
the energy of the molecule by diagonalizing the Hamiltonian
\eqref{diatomicH} using as a basis the asymptotic
electronic states
\begin{equation}
\Psi_{\mathbf{q},\mathbf{q'}}^\pm=\mathcal{N}_{\mathbf{q},\mathbf{q'}}^\pm
[
\mathrm{A}_{\mathbf{q}}(\mathbf{r}_1)\mathrm{B}_{\mathbf{q'}}(\mathbf{r}_2)\pm
\mathrm{A}_{\mathbf{q}}(\mathbf{r}_2)
\mathrm{B}_{\mathbf{q'}}(\mathbf{r}_1)],\label{defbasis}
 \end{equation}
where $\mathrm{A}_{\mathbf{p}}(\mathbf{r}_i)$
($\mathrm{B}_{\mathbf{p}}(\mathbf{r}_i)$) is a
hydrogen-like wavefunction with quantum numbers
$\mathbf{q}=(n,\ell,m)$ centered on nucleus $A (B)$, and
$\mathcal{N}_{\mathbf{q},\mathbf{q'}}^\pm=[2 (1\pm
|S_{\mathbf{q},\mathbf{q'}}|^2)]^{-1/2}$ is a normalization
factor, where $S_{\mathbf{q},\mathbf{q'}} = \int
d\mathbf{r} \, \mathrm{A}^*_\mathbf{q}(\mathbf{
r})\mathrm{B}_{\mathbf{q'}}(\mathbf{r})$ is the overlap
integral~\cite{Atkins2005}. The basis elements that are
symmetric or anti-symmetric in the coordinates have an
associated anti-symmetric (singlet) or symmetric (triplet)
function of spins respectively. Each basis
state~\eqref{defbasis} describes the two electrons being
exchanged between the two nuclei, and corresponds to an
\emph{ansatz} for a valence-bond
wavefunction~\cite{Slater1974}. The diagonal elements of
the Hamiltonian \eqref{diatomicH} using the states
\eqref{defbasis} as a basis correspond to the molecular
potentials obtained using the Heitler-London
method~\cite{Atkins2005,Coulson1952}.

The hydrogenic wavefunction of the valence electron
centered on atom $\xi$ located at $\mathbf{r}_\xi$ is well
approximated by the (unnormalized) function
\begin{equation}
\xi_\mathbf{q}(\mathbf{r})= P_\ell(\cos\theta_{\xi})\frac{\expi{m
\phi}}{a_0^{3/2}}\, \left(\frac{2
\rho_\xi}{n^*} \right)^{n^*} \mathrm{e}^{-\rho_{\xi}
/n^*} \sum_{k=0}^{k_{\mathrm{max}}}
\mathrm{b}_{k}\, \rho_\xi^{-(k+1)}.
\label{eq:defectwave}
%\mathrm{B}_{p'}(\mathbf{r})&\sim \mathrm{e}^{-r_{B}/n'} r_{B}^l L_{n'-l'-1}^{2l'+1}(r_{B}) P_l(\cos\theta_{B})
\end{equation}
Here $\rho_\xi = r_\xi/a_0$ with
$r_\xi=|\mathbf{r}-\mathbf{r}_\xi|$ the radial distance
from atomic center $\xi$, $\theta_\xi$ is the angle between
the vector $\mathbf{r}_\xi$ and the internuclear axis,
$\phi$ is the azimuthal angle, $P_\ell(x)$ is a Legendre
polynomial, and $n^*=n-\delta_\ell$ is the effective
principal quantum number, with $\delta_\ell$ a quantum
defect whose value depends on the angular momentum quantum
number $\ell$ and the atomic species~\cite{Li2003}. The
coefficients $\mathrm{b}_{k}=\mathrm{b}_{k-1}
(n^*/2k)[\ell(\ell + 1) - (n^* - k)(n^* - k + 1)]$ are
defined recursively with $\mathrm{b}_0 = 1$ and
$k_\mathrm{max}$ is an integer satisfying $n^*-\ell-1 \le
k_\mathrm{max}<n^*-\ell$~\cite{Stoly1999,Wilson1988}. The
orbital defined in Eq.~\eqref{eq:defectwave} is known as
the asymptotic form of the quantum defect wavefunction; for
$\delta_\ell=0$, it is identical to the hydrogenic
wavefunction, while for $\delta_\ell\neq 0$ it provides an
accurate description of a Rydberg electron in the mid- and
long-range. It differs significantly from the exact quantum
defect wavefunction only at short distances from the atomic
cores, which is of little consequence since the
interactions we consider here depend mainly on the outer
part of the atomic wavefunctions.

For $n>20$, the overlap between the charge-density
distributions of atoms separated by distances smaller than
the Le Roy radius (for $\mathbf{q}=\mathbf{q'}$) is
typically of the order of $S_{\mathbf{q},\mathbf{q'}}\sim
10^{-1}-10^{-2}$. The charge-density overlap is therefore
not negligible in this regime, and the approximation of the
Coulomb potential using the multipole
expansion~\cite{Marinescu1995,Marinescu1997} is not
suitable. The multipole expansion diverges quite
significantly from the Coulomb potential even in the
presence of small charge-density overlap, and so the
results obtained using this approximation below the Le Roy
radius are very uncertain~\cite{Buehler1951, Boisseau2002}.

We consequently estimate the energy of the dimer by
evaluating the expectation value of the
Hamiltonian~\eqref{diatomicH} in the states
\eqref{defbasis}, a task which requires the evaluation of a
number of integrals over atomic orbitals. These integrals
fall into two classes: the \emph{one-center} integrals,
comprising the overlap integral
$S_{\mathbf{q},\mathbf{q'}}$, the Coulomb integral
$J_{\mathbf{q},\mathbf{q'}}= j_0\int d{\mathbf r}
\,\mathrm{B}_\mathbf{q}^*(\mathbf r) (1/
r_{\!A})\mathrm{B}_{\mathbf{q'}}(\mathbf r) = j_0\int
d{\mathbf r} \,\mathrm{A}_\mathbf{q}^*(\mathbf r) (1/
r_{\!B})\mathrm{A}_{\mathbf{q'}}(\mathbf r)$, and the
charge overlap integral $K_{\mathbf{q},\mathbf{q'}}^\xi =
j_0\int d{\mathbf r} \,\mathrm{A}_\mathbf{q}^*(\mathbf
r)(1/r_{\xi})\mathrm{B}_{\mathbf{q'}}(\mathbf r)$; and the
\emph{two-center} integrals
\begin{align}
U_{\mathbf{p},\mathbf{p'}}^{\mathbf{q},\mathbf{q'}} &=
[\mathrm{A}_\mathbf{q} \mathrm{A}_{\mathbf{q'}} | \mathrm{A}_\mathbf{p} \mathrm{A}_{\mathbf{p'}} ]\quad \quad  W_{\mathbf{p},\mathbf{p'}}^{\mathbf{q},\mathbf{q'}} = [\mathrm{A}_\mathbf{q} \mathrm{B}_{\mathbf{q'}} | \mathrm{A}_\mathbf{p} \mathrm{B}_{\mathbf{p'}} ] \label{eq:exact2center}\\
V_{\mathbf{p},\mathbf{p'}}^{\mathbf{q},\mathbf{q'}} &=
[\mathrm{A}_\mathbf{q} \mathrm{A}_{\mathbf{q'}} |
\mathrm{B}_\mathbf{p} \mathrm{B}_{\mathbf{p'}} ] \quad\quad
X_{\mathbf{p},\mathbf{p'}}^{\mathbf{q},\mathbf{q'}}=[\mathrm{A}_\mathbf{q}
\mathrm{A}_{\mathbf{q'}} | \mathrm{A}_\mathbf{p}
\mathrm{B}_{\mathbf{p'}} ]\label{eq:approx2center}
\end{align}
where
\begin{align}
[\alpha \beta | \gamma \nu ]=j_0 \iint
\mathrm{d}\mathbf{r}_1\,\mathrm{d}\mathbf{r}_2 \,
\mathrm{\alpha}^*(\mathbf r_1)\mathrm{\beta}(\mathbf
r_1)\frac{1}{r_{12}}\mathrm{\gamma}^*(\mathbf
r_2)\mathrm{\nu}(\mathbf r_2).
\end{align}

The evaluation of one- and two-center molecular integrals
poses a considerable challenge for large values of the
principal quantum numbers. The results of such integrals
using direct numerical integration converge extremely
slowly, and the answers so produced can suffer from
dramatic losses of accuracy (a phenomenon known as
numerical erosion; see e.g.~\cite{Barnett2003, Rico1994}).
Following an approach suggested by M.~P.~Barnett, we use a
computer algebra-based method to generate analytical
formulae for the molecular integrals~\cite{Barnett2007,
Barnett2003,Barnett2000}. We have found this approach
advantageous for three reasons: (i) it permits the fast
evaluation of molecular integrals with high principal
quantum number to arbitrary accuracy; (ii) once the
analytical form has been found, the evaluation of an
integral for different values of the internuclear distance
is instantaneous; (iii) the analytical expressions of the
integrals can be stored and re-used at little computational
cost.

However, even with the use of symbolic calculations, the
evaluation of molecular integrals for large values of $n$
is computationally demanding. In order to make these
calculations tractable for large $n$ we restrict the value
of the projection of the electronic angular momentum along
the internuclear axis to $m=0$. This limits the range of
molecular states that can be investigated to those of
symmetry $^1\Sigma_{g}^+$ and $^3\Sigma_{u}^+$; these are
associated with the $\Psi^+_{\mathbf{q},\mathbf{q}'}$ and
$\Psi^-_{\mathbf{q},\mathbf{q}'}$ basis states
respectively. In this way we have been able to compute
molecular integrals involving wavefunctions with principal
quantum numbers up to $n=35$.

\begin{table}[t]
\begin{tabular}{|l|lcl|}
\hline Rule & Functional & & Replacement\\\hline
R1 & $\int_1^\infty \mathrm{d}x\, \mathrm{e}^{-\alpha x} x^k$ & $\rightarrow$ & $(1/\alpha^{1+k})\Gamma_{1+k}(\alpha)$\\
R2 & $\int_{-1}^1 \mathrm{d}x \,\mathrm{e}^{\beta x} x^k $ & $\rightarrow$ & $[(-\beta)^{-k}/\beta] [\Gamma_{1+k}(-\beta)-\Gamma_{1+k}(\beta)]$\\\hline
\end{tabular}
\caption{Replacement rules used to generate
analytical formulas for one-center integrals;
$\Gamma_k(x)$ is the incomplete gamma function
with $k\in \mathbb{N}_+$.} \label{rules}
\end{table}

In order to compute the molecular integrals it is
convenient to express the electron coordinates in
elliptical coordinates $(\lambda,\mu,\phi)$ through the
relations $r_\xi = (R/2)(\lambda\pm \mu)$ and
$\cos\theta_\xi= [(1\pm\lambda \mu)/(\lambda\pm\mu)]$,
where plus and minus signs apply to $\xi=A$ and $\xi=B$
respectively. The volume element is given by
$\mathrm{d}\mathbf{r}=(R/2) r_{\!A} r_{\!B}
\mathrm{d}{\lambda}\,\mathrm{d}\mu\, \mathrm{d}\phi$, where
$0 \leq \phi \leq 2\pi$, $-1 \leq \mu \leq 1$ and $1 \leq
\lambda \leq \infty$. After rounding up the powers of
$r_\xi$ in Eq.~\eqref{eq:defectwave} to the next integer
value (which only significantly affects the shape of the
wavefunction near the core) we find that the integrands of
all of the one-center integrals can be written in the form

$\mathrm{e}^{-\alpha\lambda}\mathrm{e}^{\beta \mu}
\sum_{ij} q_{i,j} \lambda^i \mu^j$, where $q_{i,j}$ are
coefficients associated with a given integral.
By applying the replacement rules defined in
Table~\ref{rules} every one-center integral can be
converted into an analytical expression with parametric
dependence on $R$.

The evaluation of symbolic expressions for two-center
integrals proceeds similarly, although many more
replacement rules are required. The general approach for
two-center integrals consists of expressing the Coulomb
potential in the form of a sum of polynomials (such as the
Legendre or Neumann expansion) before applying replacement
rules on the terms resulting from the successive
integrations over the coordinates of the first and second
valence electron. Using this method, exact formulae are
obtained for Eqs.~\eqref{eq:exact2center}, but only
approximate expressions may be found for
Eqs.~\eqref{eq:approx2center} (see
Appendix~\ref{app:rules}).

%----------------------------------------------------

\subsection{Results}
Since we envisage producing the Rydberg dimers by applying
a laser pulse to a system of ground-state atoms, we are
primarily interested in the molecular states that are most
strongly coupled to ground-state atoms by the dipole
transition operator; namely, those of $^3\Sigma_u^+$
symmetry with a high $p$-character ($\ell=1$). We will
therefore restrict our analysis to these states. However,
for values of $n\geq 10$ molecular states of $^3\Sigma_u^+$
and $^1\Sigma_g^+$ symmetry become quasi-degenerate
\cite{Marinescu1997,Boisseau2002}, and so the results
presented here apply to molecular states of either of these
symmetries.

\APSimage{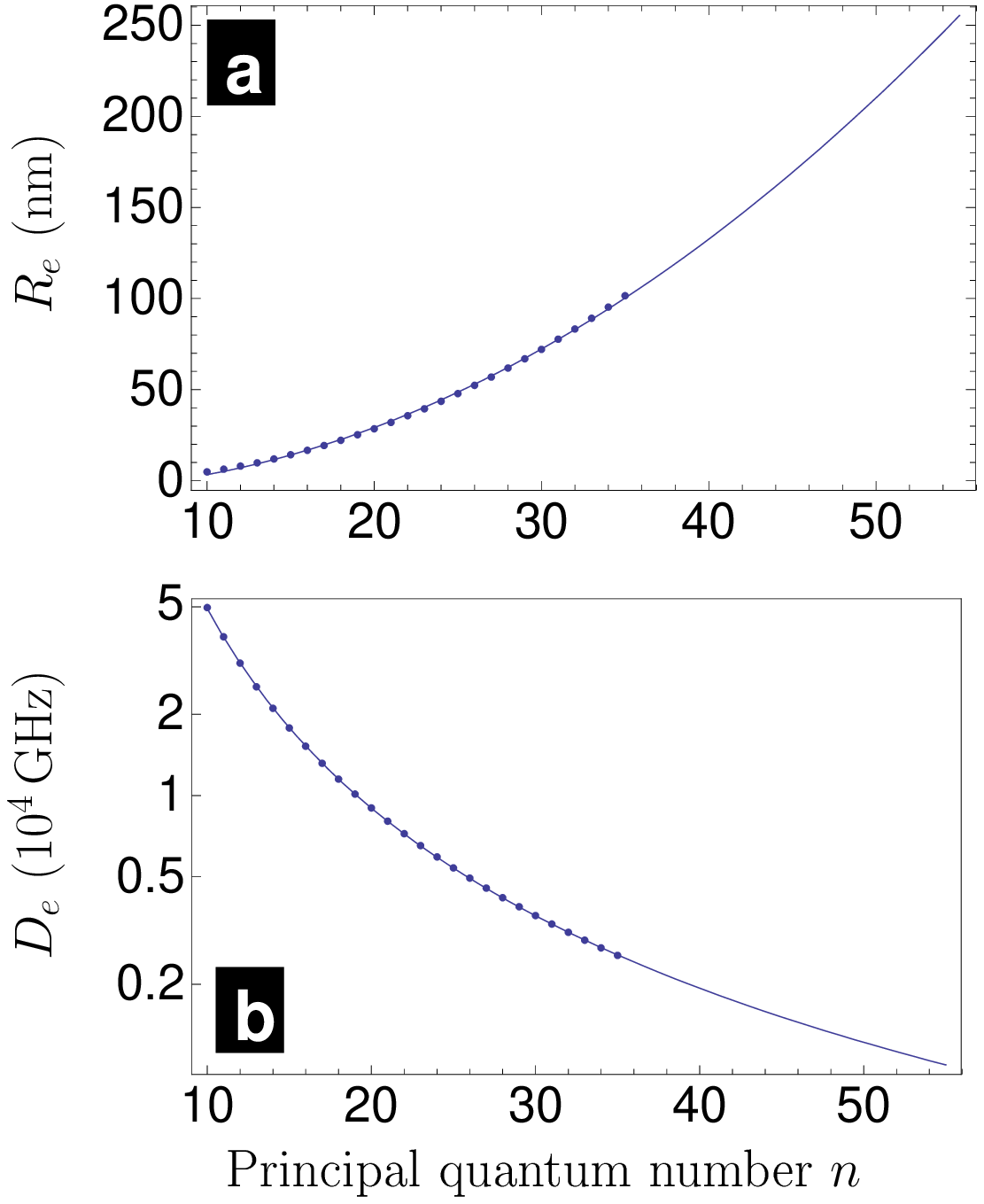}{7cm}{(color online) Equilibrium distance
$R_e$ (a) and potential depth $D_e$ of the $np+np$ states
using the Heitler-London method (b). The dots denote the values
calculated numerically, while the curves correspond to the
fitted functions mentioned in the text.}{fig:Heitler}

We have estimated the equilibrium distances $R_e$ and
potential depths $D_e = E_\mathrm{mol}(\infty)-
E_\mathrm{mol}(R_e)$ of $np+np$ molecular potentials of $^3
\Sigma_u^+$ symmetry using the Heitler-London approach for
values of the principal quantum number up to $n=35$ (see
Fig.~\ref{fig:Heitler}). We find that the equilibrium
distance of these potentials follows the relation
$R_e\simeq (1.628 n^2-100.25)\, a_0$, which is about a
factor of four smaller than the Le Roy radius for all
values of $n$~(see Fig.~\ref{fig:Heitler}a). For larger
values of the principal quantum number ($n>35$) we find
that the equilibrium distances predicted by our scaling law
are comparable to (although $\sim20\%$ smaller than) those
found by Boisseau \textit{et al.}\ in
Ref.~\cite{Boisseau2002}, who use the multipole expansion
of the Coulomb potential to examine the same potentials
below the Le Roy radius.

The depth of the molecular potential for $n=35$ is
$D_e\simeq 2000 \, \mathrm{GHz}$, approximately three
orders of magnitude larger than the potential depth of
ultra-large molecules with symmetry $^1\Pi_g-^3\Pi_u$
bonded via dipole-dipole interactions studied in Ref.~\cite{Boisseau2002}
(see Fig.~\ref{fig:Heitler}b). We find that the molecular
binding energy decreases exponentially with the value of
the principal quantum number according to the relation
$D_e=\exp (\alpha_1+\alpha_2 n^{\beta_2}+\alpha_3
n^{\beta_3}) \,\mathrm{GHz}$ where $\alpha_1=-6.53$,
$\alpha_2=-24.44$, $\alpha_3=-20.51$, $\beta_2=0.14$ and
$\beta_3=-223.45$. Our predicted potential depths differ by
up to an order of magnitude from those calculated using the
multipole expansion; as Boisseau \textit{et al.}\ point out
in Ref.~\cite{Boisseau2002}, this is probably a consequence
of the inaccuracy of the multipole expansion below the Le
Roy radius.
Although the potentials produced using our method do not
have the asymptotic $R^{-5}$ behaviour expected from
perturbation theory \cite{Marinescu1997}, this may be
corrected by carrying out a numerical integration of the
terms which were neglected during the symbolical
calculations of the molecular integrals. This correction is
feasible for small values of $n$, but very time consuming
for larger values. However, a comparison with the results
of exact numerical calculations of the same potentials up
to $n=8$ shows that our method reproduces the correct
values of $R_e$ and $D_e$ within $\sim 3 \%$.

\APSimage{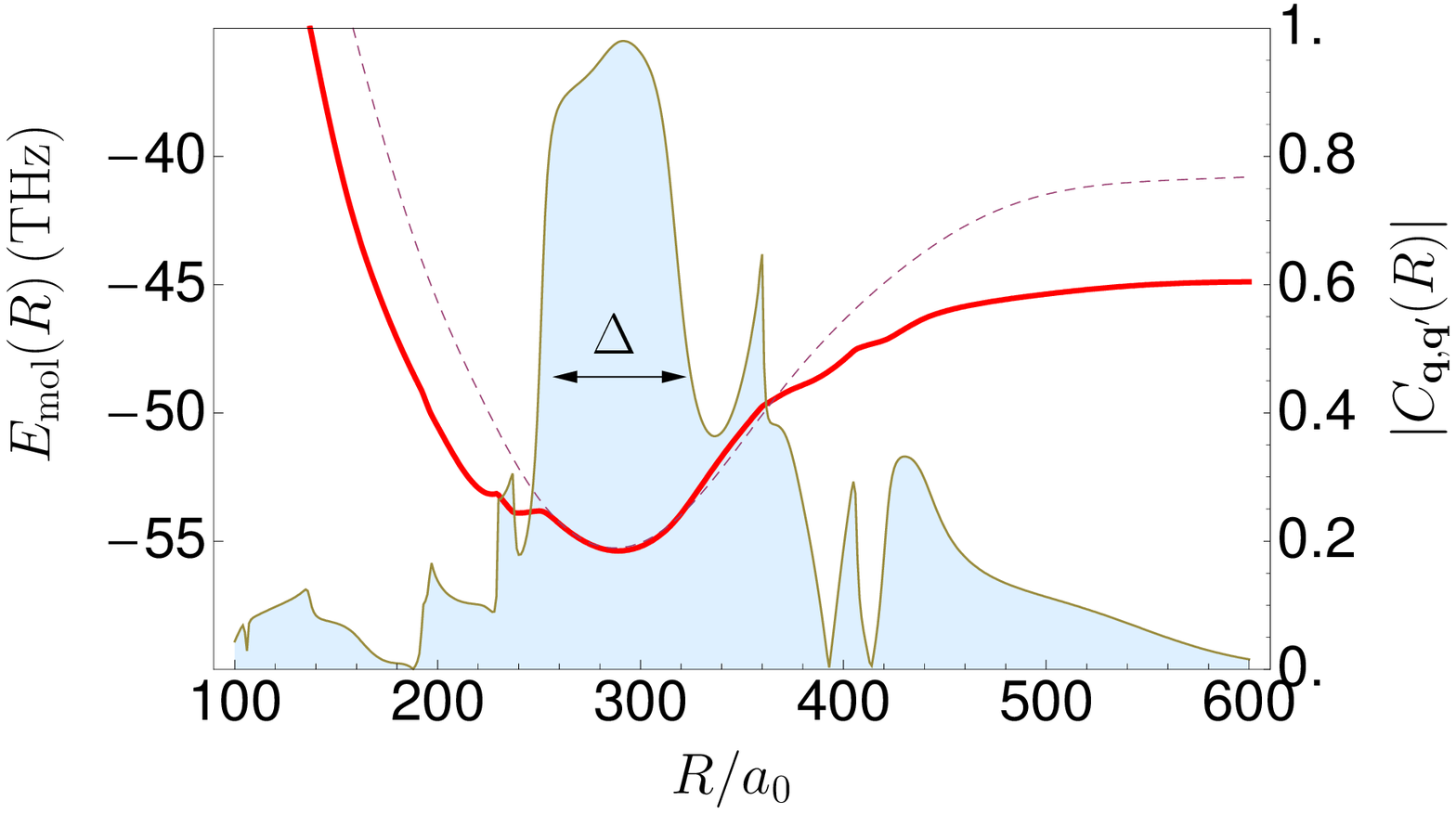}{10cm}{(color online) Left axis: The
$^3\Sigma_u$ molecular potential (solid red line)
associated with the wavefunction $\Psi_\mathrm{mol}(R)$
resulting from the diagonalization of the Hamiltonian (see
text) for $n=16$. The dashed line shows the molecular
potential for the $np+(n-1)p$ bare state. Right axis: The
contribution $C_{\mathbf{q},\mathbf{q'}}(R)$ of the
$\mathbf{q}=(16,1,0)$, $\mathbf{q}'=(15,1,0)$ basis state
to the molecular wavefunction $\Psi_\mathrm{mol}(R)$;
$\Delta$ denotes the width of the region over which this
contribution is dominant.}{fig:molecular}

Following the approach taken in Ref.~\cite{Stanojevic2008},
we diagonalized the Hamiltonian using as a basis the
molecular states~\eqref{defbasis} with a significant
coupling to the $np+n'p$ asymptote. For $n'=n-1$ and $n=16$
we used the states $(n-1)s+(n+1)s$, $(n-1)s+ns$, $(n-1)s+
(n+2-k)d$, $(n-2)d+(n-k)d$, $(n-3)d+(n-k+1)d$ and
$(n-1-k')p+(n+k)p$, with $k,k'=1,2$. The diagonalization
procedure yields eigenstates of the form
$\Psi_{\mathrm{mol}}(R)=\sum C_{\mathbf{q},\mathbf{q'}}(R)
\Psi_{\mathbf{q},\mathbf{q'}}$, allowing the determination
of the contribution of a given molecular state
$\Psi_{\mathbf{q},\mathbf{q'}}$ to each eigenstate.
Figure~\ref{fig:molecular} shows the contribution of the
asymptotic $np + (n-1)p$ state to an eigenstate
$\Psi_{\mathrm{mol}}$, and the potential curve
$E_\mathrm{mol}(R)$ corresponding to this eigenstate, for
the case of $n=16$.
In a region of width $\Delta$ about $R_e$ the molecular
potential is indistinguishable from that of the bare
$np+n'p$ state, and the associated wavefunction has a very
dominant $np+n'p$ character
($C_{\mathbf{q},\mathbf{q'}}=0.98$ with
$\mathbf{q}=(16,1,0)$ and $\mathbf{q}'=(15,1,0)$; see
Fig.~\ref{fig:molecular}). This is due to the fact that the
coupling between different states is generally weak
(off-diagonal elements of the Hamiltonian are typically
$\sim 10^{-1}-10^{-3}$ smaller than diagonal ones) and the
strongest coupling is seen between states with identical
angular momentum.

We find that the size of the region $\Delta \simeq 0.24 n^2
\, a_0$ is approximately a factor of three smaller than the
width of the molecular potential associated with the bare
$np+(n-1)p$ states (approximately given by $0.73n^2\,a_0$).
This provides an indication of the nature of the
wavefunction of the molecular states accessible by the
photoassociation of ground-state atoms initially trapped in
an optical lattice. The distance $a_\mathrm{dw}$ between
the two sides of a double well may be controlled by
altering the laser parameters. The uncertainty in the
initial position of the ground-state atoms is given by the
width of their Wannier functions~(see
Fig.~\ref{fig:setup}), which is proportional to
$\sigma=a_{\mathrm{dw}}/[\pi (\tilde{V}/E_R)^{1/4}]$, where
$\tilde{V}$ is the depth of the double-well and $E_R$ the
recoil energy~\cite{jaksch1}. By setting
$a_\mathrm{dw}=R_e$ and considering the scaling of the
width of the $np+(n-1)p$ molecular potentials, we find that
$2 \sigma\lesssim \Delta$ for lattice potential depths of
the order of $\tilde{V}=30$--$40\,E_R$. This suggests that
using an optical lattice to enforce the initial position of
the atoms before applying the photoassociation pulse might
offer advantages in providing access to molecular states
with a very dominant $np+n'p$ character.

We have obtained an order-of-magnitude estimate of the
lifetime of the photoassociated molecules by assuming for
simplicity that the dominant decay mode is radiative
dissociation into a pair of $^{87}\mathrm{Rb}$ atoms. We
neglect the rotational fine structure of the energy levels
and any bound-bound decay channels and consider a
transition between a bound state of vibrational quantum
number $\nu'$ and energy $E_{\nu'}$ and a free continuum
state of wavenumber $k''$ and energy
$E_{k''}=\hbar^2k''^2/(2\mu)$, where $\mu =
m_{\mathrm{Rb}}/2$ is the reduced mass of the molecule. The
Einstein $A$ coefficient is given by
\begin{align}\label{EinsteinACoeff}
A_{\nu' k''} =&\frac{32\pi^3}{3\varepsilon_0 h^5 c^3}
\sqrt{\frac{2\mu}{E_{k''}}}(E_{\nu'}-E_{k''})^3 \nonumber\\
&
\times\left|\int\left[\psi^{\mathrm{vib}}_{\nu'}(R)\right]^*D(R)\psi^\mathrm{vib}_{k''}(R)\,dR\right|^2
\quad \mathrm{ J^{-1}\,s^{-1}}
\end{align}
while the radiative lifetime of a single bound vibrational
level is given by $\tau_{\nu'} = A _{\nu'}^{-1}$, where
\begin{align}
  A_{\nu'} = \int_0^{\infty} A_{\nu'
  k''}\, dE_{k''}.
\end{align}
Here $h = 2\pi\hbar$ is the Planck constant, $c$ is the
speed of light, $\psi^{\mathrm{vib}}_{\nu'}$ and
$\psi^\mathrm{vib}_{k''}$ are vibrational wave functions
for the discrete and continuum states respectively, and the
dipole moment $D(R)$ is given by
\begin{align}
D(R)=-e \iint \mathrm{d}\mathbf{r}_1
\mathrm{d}\mathbf{r}_2\,
[\Psi_{\mathbf{q}_i,\mathbf{q}_i}^\pm]^*(z_1+z_2)
\Psi_{\mathbf{q}_f,\mathbf{q}_f}^\pm
\end{align}
where $z_1$ and $z_2$ are the $z$-coordinates of the
electrons and $\mathbf{q}_i$ and $\mathbf{q}_f $ denote the
quantum numbers of the initial and final states. For large
internuclear distances the continuum wave function
$\psi^\mathrm{vib}_{k''}$ has the asymptotic form
\begin{align}\label{eqn:scatteringchi}
  \psi^\mathrm{vib}_{k''} \sim \sin(k''R+\eta)
\end{align}
where $\eta$ is the scattering phase shift, which at low
energies is given by $\eta = -k''a_s$ with $a_s$ the
$s$-wave scattering length. In the initial molecular state
we approximate the true bound vibrational wavefunctions
$\psi^\mathrm{vib}_{\nu'}(R)$ by the eigenstates of a
harmonic approximation to the molecular potential centered
about $R = R_e$. Due to the asymmetric nature of the
molecular potential curves, this approximation becomes
progressively worse as $\nu'$ increases. However, we expect
that the ability to impose an interatomic spacing close to
the equilibrium distance of the targeted molecular state
before applying the photoassociation pulse will provide a
measure of control over the range of vibrational levels
occupied by the molecules, thus making the lowest
vibrational levels the most relevant.

Although there are a huge number of final states to which
the molecule could decay, the dependence of the decay rate
given by Eq.~\eqref{EinsteinACoeff} on
$(E_{\nu'}-E_{k''})^3$ favors transitions that involve the
emission of a high-energy photon. We therefore consider
only transitions that finish within the continuum above the
5$s$--5$s$ potential curve, neglecting all other decay
channels. The radiative lifetimes thus calculated for the
ground vibrational state $\nu'=0$ are shown in
Fig.~\ref{fig:lifetimes}; the lifetimes of higher
vibrational levels (up to $\nu'=10$) are the same to within
$\sim5\%$. Also plotted for comparison purposes is the
radiative lifetime $\tau = \tau_0 n^3$ of a free Rydberg
atom, where $\tau_0 = 1.4$~ns for
$^\mathrm{87}$Rb~\cite{Gallagher1994}. Our calculations
indicate that for $n\gtrsim17$ the Rydberg dimers are more
stable than the free atoms. For the highest molecular state
considered ($n=35$) we find lifetimes of the order of a few
milliseconds, indicating that even if contributions from
neglected decay channels were to reduce this lifetime by
several orders of magnitude, the system could still be
successfully interrogated and characterised by short
(nanosecond to picosecond) laser pulses. To avoid possible
stimulated emission contributions to the radiative decay
process, the laser fields forming the optical lattice could
be switched off as the Rydberg excitation pulse is applied.
The subsequent free expansion of the atoms would not limit
the window of time within which the system may be
characterized, since the atoms are expected to remain
localized within a few lattice spacings for a time of the
order of 10 $\mu$s \cite{Toth2008}.

\APSimage{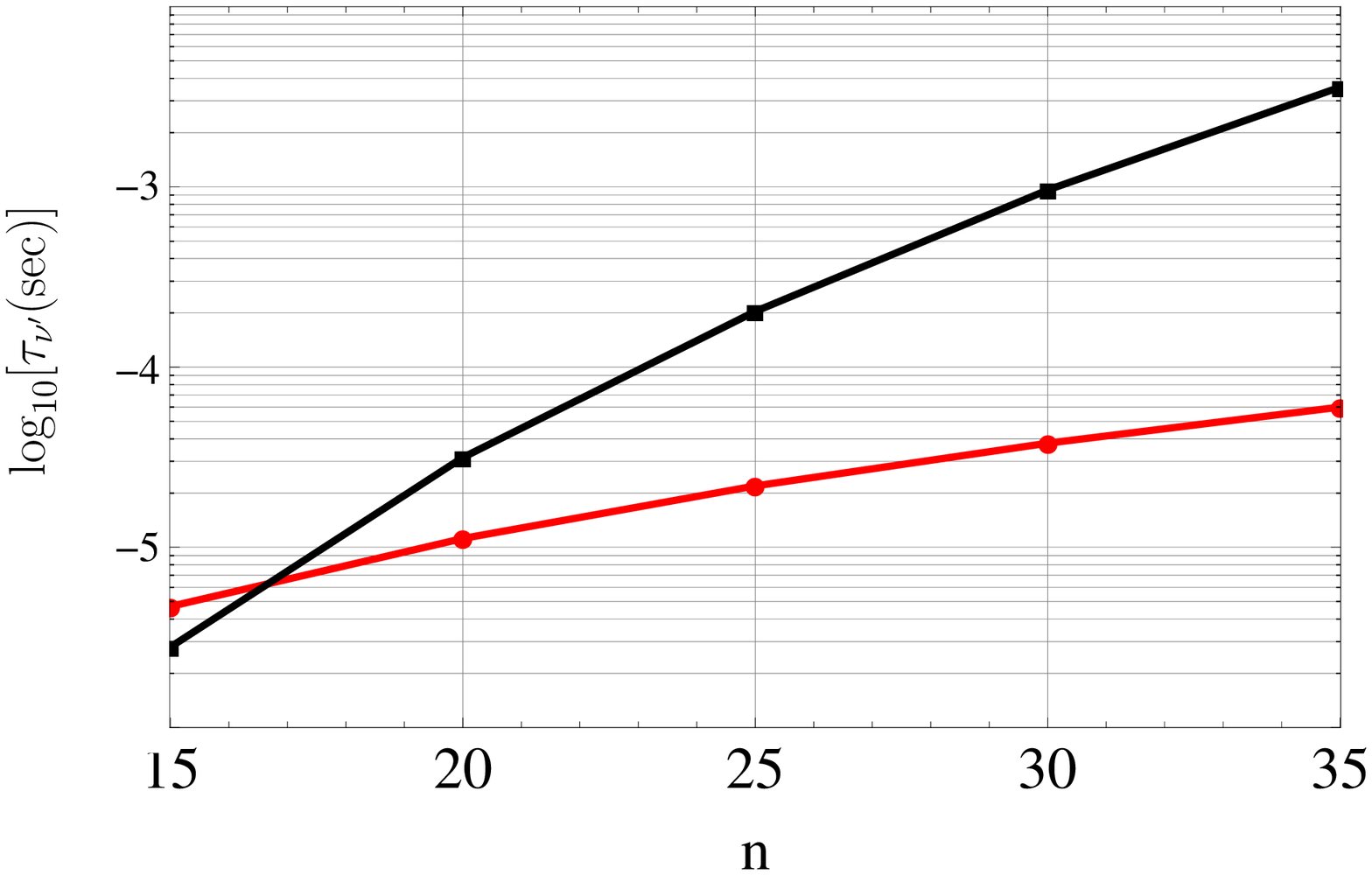}{8cm}{(color online) Radiative lifetimes for the
$\nu'=0$ vibrational level in molecular excited states with
principal quantum numbers $n$ = 15--35 (black line). The
radiative lifetime $\tau = \tau_0 n^3$ of a free Rydberg
atom is plotted for comparison purposes (red line).}{fig:lifetimes}

%----------------------------------------------------
\section{Interactions between highly excited electrons in a lattice\label{sec:simul}}

In this section we consider the situation where ground
state atoms trapped in a regular optical lattice are
collectively excited to a given Rydberg state such that the
charge-density distributions of neighboring atoms overlap.
With overlapping charge-density distributions between
neighboring atoms, the valence electrons will tunnel
between lattice sites and interact with each other,
mimicking the behaviour of electrons in metals. Such a
system may have interesting applications for the quantum
simulation of electrons in lattice systems. Assuming that the band structure of such a system is well described by a tight-binding model, we will find that using the hopping rate calculated in the next section that the Fermi
temperature of a half-filled lattice is $T_F\sim 10^{-3}\,\mathrm{K}$, some
5 orders of magnitude larger than typical temperatures
achieved in experiments with ultracold atoms. While the
efficient population transfer of atoms to Rydberg states is
experimentally very challenging and goes beyond the scope
of this paper (see e.g.~Ref.~\cite{Cubel2005}), this figure
implies that after the excitation step, the system temperature could still remain well within the Fermi degeneracy regime $(T\ll
T_F)$. This would enable the observation of interesting
many-body phenomena currently inaccessible to other
experimental setups involving fermionic atoms where the ratio $T/T_F\approx 0.25$~(see
e.g.~Ref.~\cite{Kohn2005, Lewenstein2006}). In this
section, we use the tools developed previously to evaluate the typical interaction parameters and the hopping rate of such a
system, and investigate their dependence on the interatomic
spacing fixed by the lattice.

%----------------------------------------------------
\subsection{Model}
In this section we assume that ground-state atoms trapped
in an optical lattice can be collectively excited to a
given Rydberg state. We assume that the dynamics of the valence electrons in the lattice is restricted to only one spatial mode per site $i$ that corresponds to an orbital $\phi_i(\mathbf{r})=\braket{\mathbf{r}}{\phi_i}$ associated with a given Rydberg state localized at site $i$ with two spin orientation. The Hamiltonian of the system is then given by
\begin{equation} \op{H}=-\sum_{i,j,\sigma} t_{ij}
\,\cre{c}_{i\sigma}
\des{c}_{j\sigma}+\frac{1}{2}\sum_{\substack{i,j,k,l\\
\sigma,\sigma'}}V(i,j,k,l)
\cre{c}_{i,\sigma}\cre{c}_{j,\sigma'}\des{c}_{k,\sigma'}\des{c}_{l,\sigma},\label{genHubbard}
\end{equation}
where $\cre{c}_{i\sigma}$ is a creation operator associated with the orbital $\phi_i(\mathbf{r})$ and spin $\sigma$, $t_{ij}$ describes
the hopping of a particles between sites $i$ and $j$; and
the inter-site interactions are given by
\begin{equation}
 V(i,j,k,l)= \iint \mathrm{d}r\,\mathrm{d}r' \phi_i^*(r)\phi_j^*(r') V_{\mathrm{ee}}(r-r')\phi_k(r)\phi_l(r'),\label{definteraction}
\end{equation}
where $V_{\mathrm{ee}}(r-r')$ is the interelectronic
potential. We assume here that next-nearest-neighbour
hopping can be neglected, and consider only
nearest-neighbour interactions. In this situation the
hopping term $t=t_{i,i+1}$ consists of the kinetic energy
and Coulomb potential of neighboring ion cores
\begin{equation}
 t=\bra{\phi_i} \left(-\frac{\hbar^2}{2 M} \nabla^2 - \frac{j_0}{\left| \mathbf{r}-\mathbf{R}_i \right|}- \frac{j_0}{\left| \mathbf{r}-\mathbf{R}_{i+1} \right|}\right)\ket{\phi_{i+1}},\label{defhopping}
\end{equation}
and the only two-electron interaction terms taken
into account are $\mathcal{U}=V(i,i,i,i)$,
$\mathcal{X}=V(i+1,i,i,i)$,
$\mathcal{V}=V(i,i+1,i+1,i)$ and
$\mathcal{W}=V(i,i+1,i,i+1)$~\cite{Appel1993}.

\APSimage{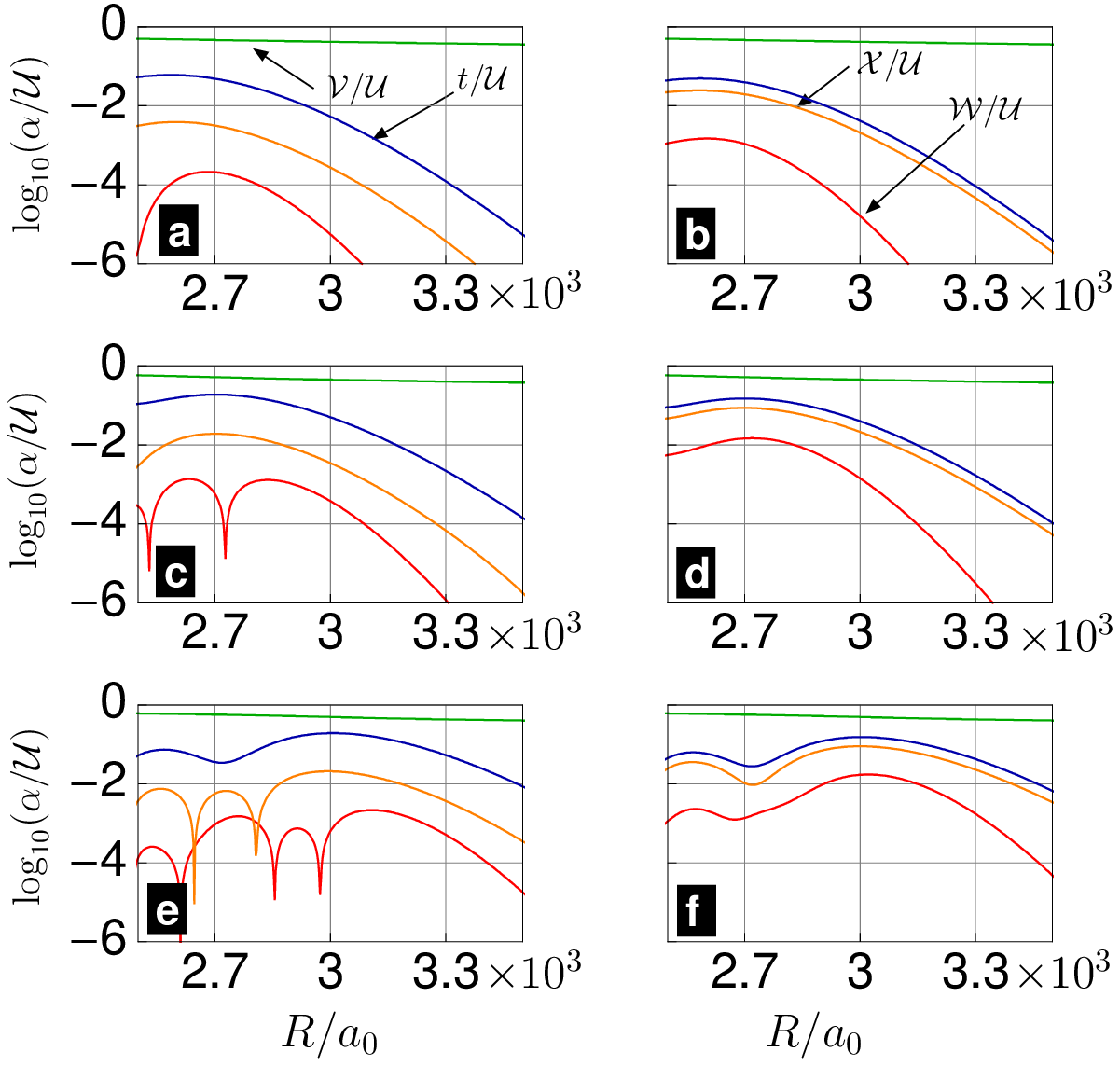}{9cm}{(color online) Relative values of
the interaction parameters ($\alpha=\mathcal{V},t,\mathcal{X},\mathcal{W}$) of a Hubbard model of Rydberg
electrons in a state with quantum numbers
$\mathbf{q}=(30,\ell,0)$. (a) $\ell=0$ (b) $\ell=0$,
setting $S_{\mathbf{q},\mathbf{q}}=0$; (c) $\ell=1$ (d)
$\ell=1$, setting $S_{\mathbf{q},\mathbf{q}}=0$; (e)
$\ell=2$ (f) $\ell=2$, setting
$S_{\mathbf{q},\mathbf{q}}=0$. From top to bottom we have
$\mathcal{V}/\mathcal{U}$ (green), $t/\mathcal{U}$ (blue),
$\mathcal{X}/\mathcal{U}$ (orange), and
$\mathcal{W}/\mathcal{U}$ (red). }{fig:fermiparam}

Electronic orbitals centered at different lattice sites are
often considered to be orthogonal and equated to the
Wannier functions of the crystal~\cite{Ashcroft1976}.
However, a better definition of the Wannier function
centered at site $i$ in terms of electronic orbitals is
given by
\begin{equation}
 \phi_i= \psi_i - \frac{S_i \psi_{i+1}+S_{i-1}\psi_{i-1}}{2},\label{Wannier}
\end{equation}
where $\psi_i$ is a normalized electronic wavefunction with
quantum numbers centered at site $i$, and $S_i$ is the
overlap between the orbitals of site $i$ and
$i+1$~\cite{Painelli1989}. In the regime where the terms
proportional to $S_i^2$ can be neglected, the functions
$\phi_i$ form an orthonormal basis set. By inserting the
definition~\eqref{Wannier} into Eqs.~\eqref{definteraction}
and~\eqref{defhopping} and assuming that the site orbitals
$\psi_i=\Psi_\mathbf{q}(\mathbf{r}-\mathbf{R}_i)$
correspond to hydrogenic wavefunctions with quantum numbers
$\mathbf{q}$, the parameters of the
Hamiltonian~\eqref{genHubbard} can be expressed in terms of
two-center molecular integrals as
\begin{align}
 \mathcal{U}&=U_{\mathbf{q},\mathbf{q}}^{\mathbf{q},\mathbf{q}}-4 S_{\mathbf{q},\mathbf{q}} X_{\mathbf{q},\mathbf{q}}^{\mathbf{q},\mathbf{q}}\,, \nonumber\\
 t&=(E_\mathbf{q}+j_0/R)S_{\mathbf{q},\mathbf{q}}-2 K_{\mathbf{q},\mathbf{q}}\nonumber\\
\mathcal{V}&=V_{\mathbf{q},\mathbf{q}}^{\mathbf{q},\mathbf{q}}-2 S_{\mathbf{q},\mathbf{q}} X_{\mathbf{q},\mathbf{q}}^{\mathbf{q},\mathbf{q}}\, ,\, \mathcal{W}=W_{\mathbf{q},\mathbf{q}}^{\mathbf{q},\mathbf{q}}-2 S_{\mathbf{q},\mathbf{q}} X_{\mathbf{q},\mathbf{q}}^{\mathbf{q},\mathbf{q}},\nonumber\\
\mathcal{X}&=X_{\mathbf{q},\mathbf{q}}^{\mathbf{q},\mathbf{q}}- S_{\mathbf{q},\mathbf{q}}[ W_{\mathbf{q},\mathbf{q}}^{\mathbf{q},\mathbf{q}}+\frac{1}{2}(U_{\mathbf{q},\mathbf{q}}^{\mathbf{q},\mathbf{q}}+V_{\mathbf{q},\mathbf{q}}^{\mathbf{q},\mathbf{q}})],\label{fermiparam}
\end{align}
where $K_{\mathbf{q},\mathbf{q}}=K_{\mathbf{q},\mathbf{q}}^A=K_{\mathbf{q},\mathbf{q}}^B$, $\mathcal{U}$ is the on-site interaction,
$\mathcal{V}$ and $\mathcal{W}$ are off-diagonal repulsion
terms, and $\mathcal{X}$ is sometimes called the
density-dependent hopping or enhanced hopping
rate~\cite{Marsiglio1990}. The absolute value of the ratios
between different interaction parameters for atoms in
$np_z$, $ns$ and $nd_{z^2}$ states ($n=30$) are shown
in~Fig.~\ref{fig:fermiparam}, where we have plotted the
parameters in Eq.~\eqref{fermiparam} obtained with $S_i=0$, that is,
with Wannier functions represented by bare atomic orbitals.
The orthogonalization procedure (i.e. replacing $\psi_i$ by
$\phi_i$) mainly affects the smallest parameters
$\mathcal{X}$ and $\mathcal{W}$ in the regions where the
value of $S_i^2$ is not small enough for the Wannier
functions to be effectively orthogonal.

The interaction parameters calculated above correspond to
the intermediate screening case, that is
$\mathcal{U}>\mathcal{V}\gg \mathcal{W}$ and
$\mathcal{X}\simeq \kappa \mathcal{J}$, where $\kappa$ is a
constant~\cite{Gammel1988,Baeriswyl1988}. The ratios
between the parameters are similar to those found in
realistic systems, e.g.~between $p$-electrons in conjugated
polymers~\cite{Gammel1988}. For different values
of the angular momentum, these ratios differ mainly in the way they
behave at large distances. The hopping rates $t$ are shown
in Fig.~\ref{fig:hopping}, and, for the quantum numbers
considered, vary between $10^2$ and $10^{-4}\,\mathrm{GHz}$
for intersite distances between $150$ and $250\,
\mathrm{nm}$
respectively. 

\APSimage{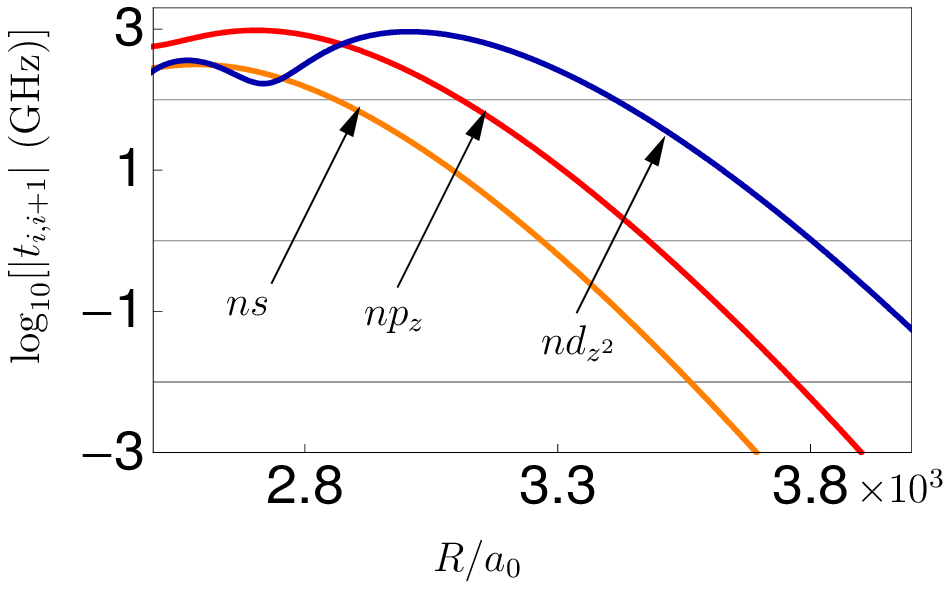}{8cm}{(color online) Hopping rate in the
$z$-direction between sites separated by a distance $R$ for
electrons in $ns$, $np_z$, $nd_{z^2}$ states
($n=30$).}{fig:hopping}

% %(* Rem for n=40, radiative lifetime is 89 microsec *)
%----------------------------------------------------
\section{Conclusion\label{sec:conclusion}}
We showed that optical lattices forming arrays of
double-well potentials may be exploited to selectively
photoassociate pairs of atoms to molecular states with
binding energies of the order of $10^3\, \mathrm{GHz}$, far
larger than those of long-range molecules stabilized by
dipole-dipole forces. These molecular states are expected
to have equilibrium distances of the order of the typical
lattice spacings and lifetimes several orders of magnitude
larger than the timescales required to interrogate and
characterize them by way of short laser pulses.

We considered the possibility of collectively exciting
ground-state atoms trapped in a regular lattice to a given
Rydberg level such that the charge-density distributions of
atoms located in neighboring sites overlap. Assuming that
such a system could be realized, we calculated the typical
interaction parameters between the Rydberg electrons and the hopping rate between sites. With
a temperature well below the Fermi temperature and tunable
interaction parameters, such systems might offer an interesting
alternative approach to the simulation of Fermi systems.
For instance, the ability to change the lattice spacing
such that $v=\mathcal{U}/\mathcal{V}>\frac{1}{2}$ or
$v<\frac{1}{2}$ would offer the possibility of engineering
a charge-density wave (CDW) or a spin-density wave (SDW),
respectively. The preparation of a SDW state could even be
facilitated by using spin-changing collisions between
ultracold $^{87}\mathrm{Rb}$ ground-state atoms to produce
a N\'{e}el-like state $\ket{\uparrow\downarrow \uparrow
\downarrow \dots}$ that has a spin arrangement identical to
that of the SDW ground state along one spatial
direction~\cite{Trotzky2008}. If realized, such a system
would allow the observation of a phase transition between
the SDW and CDW phases, by, for example, the measurement of
the zero-frequency SDW susceptibility~\footnote{The SDW
susceptibility and the structure factor are defined
respectively by $\chi(q)=(1/N)\int_0^{1/T}
{\mathrm{d}}\tau\sum_{i,j} \kappa_i(\tau)\kappa_j(0)$ and
$S(q)=(1/N) \sum_{i,j} \expi{q(R_i-R_j)}\langle n_i n_j
\rangle$, where
$\kappa_i(t)=n_{i\uparrow}(t)-n_{i\downarrow}$ and
$n_{i,\sigma}=\cre{c}_{i,\sigma}\des{c}_{i,\sigma}$.} or
the CDW structure factor, both of which diverge linearly in
their respective phase~\cite{Rey2005}. Further, the versatility of optical lattice setups may allow the excitation of Rydberg atoms with a given angular momentum, which potentially enables exotic quantum phase
transitions to be engineered: indeed, the value of the
electronic angular momentum not only influences the
relative values of the system interaction parameters, but
also in some cases their signs~(see
e.g.~Refs.~\cite{Kaxiras2002,Elfimov2002,Hirsch1993}). This
feature is particularly interesting, as it is the mechanism
behind e.g.~the emergence of non-trivial and rich phase
diagrams in doped cuprates (see
e.g.~Ref.~\cite{Duffy1995,Racz2006}). Finally, if in addition the electron density could be controlled when producing one of these
states, it would provide the exciting possibility of turning
the state of the electrons into a superconductor, as
happens in the case of doping-induced superconductivity in
cuprates~\cite{Sachdev2002}. The parameters calculated in this work apply to models where the dynamics of the electrons is restricted to one mode per site. 

In future works, it would be interesting to determine the conditions for this assumption to be valid, and whether these conditions can be engineered with current technology. Results in this direction would allow excitation schemes to be devised, and also the determination of the control available to tune the density of electrons in the lattice. Also, dynamical calculations aiming at characterizing the lifetime of the lattice configuration after the atoms have been excited to Rydberg levels would allow the evaluation of the time-scale available to interrogate the system. The exact calculation of this lifetime is an open problem whose solution is beyond the scope of the present paper. However, the dynamics of our system happens on a time scale of a few hundreds of picoseconds. Techniques for probing electron dynamics on the femto-second time scale exist in condensed matter systems (see e.g. Ref.~\cite{Cavalleri2005})). These methods might form the basis for resolving the motion of electrons in a Rydberg gas on pico- to femto-second timescales in future experiments. The use of our scheme for the purpose of quantum simulation requires the lifetime of the lattice configuration to be longer than the typical timescales of the dynamics. As mentioned by Mourachko {\it et al.} in Ref.~\cite{Mourachko2004}, the interatomic spacing ($\sim \mu$m) between Rydberg atoms in a frozen gas varies very little ($\sim 3$\%) over a period of time of the order of 1 $\mu$s, some three orders of magnitude larger than the typical hopping times present in the system we have proposed. Also, as already mentioned by Li {\it et al.} in Ref.~\cite{Li2005} fixing the initial positions of the atoms will help reducing the motion of the ion cores. We therefore believe that Rydberg gases created from ultracold atoms in an optical lattice provide a promising route towards the direct quantum simulation of interacting fermi systems.

%----------------------------------------------------

\section*{Acknowledgements}
B.V.\ would like to thank A.\ Nunnenkamp and
Prof.\ M.\ Child for helpful discussions at the
beginning of this project. This work was
supported by the EU through the STREP project
OLAQUI, and by the University of Oxford through
the Clarendon Fund~(S.T.). B.V.\ acknowledges
partial financial support from Merton College
(Oxford, UK) through the Simms Bursary.

%----------------------------------------------------
\appendix

\section{Replacement rules for the evaluation of two-center molecular integrals with $m=0$.}
\label{app:rules} Here we outline the methods we have used
to solve molecular integrals using symbolic replacements.
We also provide all the necessary relations to implement
the method for $m=0$.

The first step towards the evaluation of these integrals is
to express the Coulomb potential in terms of a series of
Legendre functions as
\begin{align}
 \frac{1}{r_{12}}&= \sum_{k=0}^\infty \sum_{m=-k}^{k} \frac{(k-|m|)!}{(k+|m|)!} \frac{r(a)^k}{r(b)^{k+1}} \times \nonumber \\
& \qquad P_k^{|m|}(\cos \theta_1) P_k^{|m|}(\cos \theta_2) \expi{m(\phi_1 - \phi_2)},\label{eq:coulombLeg}
\end{align}
where $r_{12}$ is the distance between two points with
spherical coordinates $(r_i, \theta_i, \phi_i)$,
$P_k^{|m|}(x)$ are the associated Legendre functions [we
use the notation $P_k^{0}(x)=P_k(x)$], and $r(a)$, $r(b)$
are the smaller and larger of the quantities $r_1$ and
$r_2$; or using the von Neumann expansion
\begin{align}
  \frac{1}{r_{12}} &= \frac{2}{R}\sum_{k=0}^\infty \sum_{m=-k}^{k} (-1)^m (2k+1) \left[\frac{(k-|m|)!}{(k+|m|)!}\right]^2 \times\nonumber \\
&\qquad P_k^{|m|}[\lambda(a)] Q_k^{|m|}[\lambda(b)] P_k^{|m|}(\mu_1)P_k^{|m|}(\mu_2) \expi{m(\phi_1 - \phi_2)},\label{eq:coulombVon}
\end{align}
where in this case the two points are expressed in
elliptical coordinates $(\lambda_i, \mu_i, \phi_i)$,
$Q_k^{|m|}(x)$ are associated Legendre functions of the
second kind, and $\lambda(a)$ is the lesser and
$\lambda(b)$ the greater of $\lambda_1$ and
$\lambda_2$~(see e.g. Ref.~\cite{Slater1974}).

Since the integral $\int_0^{2\pi}\expi{\nu \phi}
{\mathrm{d}}\phi$ vanishes if $\nu$ is an integer different
from zero, setting the quantum number $m=0$ considerably
simplifies the evaluation of the
integrals~\eqref{eq:exact2center}
and~\eqref{eq:approx2center} using the
relations~\eqref{eq:coulombLeg} and~\eqref{eq:coulombVon}.

Every two-center molecular integral apart from
$W^{\mathbf{q},\mathbf{q'}}_{\mathbf{p},\mathbf{p'}}$ can
be evaluated using Eq.~\eqref{eq:coulombLeg}. The angular
part of these integrals will be non-zero for only a few
terms; for instance, for $\mathbf{q}=\mathbf{q'}=(n,0,0)$,
the angular part is non-zero for $k=0$, and for
$\mathbf{q}=\mathbf{q'}=(n,1,0)$ for $k=0,2$. It may be
simplified by using the formula for the product of surface
harmonics:
\begin{equation}
 P_{\ell_1}(\cos \theta)P_{\ell_2}(\cos \theta)= \sum_{j=|\ell_1 -\ell_2|}^{\ell_1 +\ell_2} [C^{\ell_1,\ell_2}_{j}]^2 \, P_j(\cos \theta),
\end{equation}
where $C^{\ell_1,\ell_2}_{j}=C(\ell_1,\ell_2,j;0,0,0)$ are
Clebsch-Gordan coefficients~\cite{Barnett2000}. The
integration of the radial part over the coordinates of the
first electron necessitates the evaluation of integrals of
the form
\begin{align}
 \int_0^\infty {\mathrm{d}}r_1\, r_1^2 \frac{r(a)^k}{r(b)^{k+1}} a_\mathbf{q}(r_1)a_{\mathbf{q'}}(r_1)=I_{\mathbf{qq'},k}^{+}(r_2)+I_{\mathbf{qq'},k}^{-}(r_2),
\end{align}
where $a_\mathbf{q}(r)$ is the radial part of the
wavefunction~\eqref{eq:defectwave}, and $r_i$ is the
coordinate of the $i^{\mathrm{th}}$ electron. The
functionals $I_{\mathbf{qq'},k}^{\pm}(r_2)$ are defined as
\begin{align}
I_{\mathbf{qq'},k}^{-}(r_2)= (1/r_2^{k+1}) \int_0^{r_2}
{\mathrm{d}}r_1\, r_1^2 r_1^k
a_\mathbf{q}(r_1)a_{\mathbf{q'}}(r_1)\label{eq:Iminus}
\end{align}
and
\begin{equation}
I_{\mathbf{qq'},k}^{+}(r_2)= r_2^{k} \int_{r_2}^\infty
{\mathrm{d}}r_1\, r_1^2 r_1^{-k-1}
a_\mathbf{q}(r_1)a_{\mathbf{q'}}(r_1).\label{eq:Iplus}
\end{equation}
For $k=0$ both \eqref{eq:Iminus} and~\eqref{eq:Iplus} can
be solved using the rules
\begin{align}
 \mathrm{R}3:&\int_0^r {\mathrm{d}}r\, {\mathrm{e}}^{-a r} r^k &\rightarrow&\quad [k! - \Gamma_{k+1}(a r)]/a^{k+1}\nonumber\\
 \mathrm{R}4:&\int_r^\infty {\mathrm{d}}r\, {\mathrm{e}}^{-a r} r^k &\rightarrow&\quad \Gamma_{k+1}(a r)/a^{k+1}\label{eq:rel1}
\end{align}
If the sum of Eq.~\eqref{eq:coulombLeg} contains only $k=0$
terms, the symbolic form of the integral is obtained by
replacing the spherical coordinates of the remaining
electron by elliptical coordinates and using the
replacement rules of Eqs.~\eqref{eq:rel1} and
Table~\ref{rules}. In integrals requiring terms associated
with $k>0$ in the sum~\eqref{eq:coulombLeg}, the
functionals associated with $k=0$
dominate~\cite{Appel1993}. For higher values of $k$, the
functional~\eqref{eq:Iplus} can always be solved, and as an
approximation we have dropped the terms
in~\eqref{eq:Iminus} that could not be solved analytically
using the integration rules mentioned above. 

Because it involves associated Legendre functions of the
second kind, the evaluation of $W^{\mathbf{q},\mathbf{q'}}_{\mathbf{p},\mathbf{p'}}$ is more
problematic. We used the Rodrigues formula to expand these
functions in terms of sums of Legendre polynomials and
logarithms, and then solved the polynomial part by applying
the replacement rules mentioned above. Solving the
logarithmic part requires the integration over $\log(1 \pm
x)$ for the first set of coordinates, and then exponential
integral functions for the second set.

%\bibliography{/home/vaucher/bibliography/main}

\end{document}